\documentclass[%
 aip,
 pof,%
 amsmath,amssymb,
reprint,%
]{revtex4-1}
\usepackage{CJK}
\usepackage{graphicx}
\usepackage{dcolumn}
\usepackage{bm}
\usepackage{color}
\usepackage{hyperref}
\usepackage{upgreek}

\newcommand{\Nu}{N\!u}%
\usepackage{makecell}
\begin{document}
\preprint{AIP/123-QED}

\title{Conjugate Heat Transfer Effects on Bubble Growth During Flow Boiling Heat Transfer in Microchannels}

\author{Odumuyiwa A. Odumosu}
\affiliation{State Key Laboratory of Engines, Tianjin University, Tianjin, 300350, China.}%
\author{Hongying Li}
\affiliation{School of Mechanical and Aerospace Engineering, Nanyang Technological University, 50 Nanyang Avenue, Singapore 639798, Republic of Singapore.}
\author{Tianyou Wang}
\affiliation{State Key Laboratory of Engines, Tianjin University, Tianjin, 300350, China.}%
\affiliation{National Industry-Education Platform of Energy Storage, Tianjin University, Tianjin, 300350, China}
\author{Zhizhao Che}
\email{chezhizhao@tju.edu.cn}
 \affiliation{State Key Laboratory of Engines, Tianjin University, Tianjin, 300350, China.}
 \affiliation{National Industry-Education Platform of Energy Storage, Tianjin University, Tianjin, 300350, China}
\date{\today}

\begin{abstract}
Flow boiling in microchannel heat sinks is an efficient way to dissipate high heat flux by utilizing the large surface-to-volume ratio and high latent heat. Previous studies of boiling heat transfer in microchannels mainly consider the fluid flow in channels only, but often neglect the conjugate effects of the heat conduction in the solid wall, which becomes important for microchannels because of the comparable sizes of the flow channel and the solid wall. In the present study, the effects of conjugate heat transfer on bubble growth during flow boiling in microchannels are examined by numerical simulation. The results indicate that the bubble growth is non-uniform for different bottom wall thicknesses or different solid materials even with the same heat flux at the wall. As the bottom wall thickness increases, the bubble growth rate increases because of the heat conduction in the solid wall along the channel direction. The increased bubble size also increases the perturbation to the flow field, and enhances the thermal convection between the fluid and the wall. For different solid materials, the high-thermal-diffusivity material possesses a higher heat transfer performance because of the quick diffusion of thermal energy from the heat source to the solid-fluid interface.
\end{abstract}


\maketitle
\onecolumngrid
\section{Introduction}\label{sec:1}
The increasing miniaturization of electronic devices has necessitated the development of high-heat-flux removal techniques \cite{ghani17, leong17}. Flow boiling in microchannels is an efficient way to dissipate high heat flux by utilizing the high surface-to-volume ratio and excellent cooling performance. Apart from electronic devices, other applications of microchannel heat sinks include lasers, batteries, electric vehicles, avionics, photovoltaics, solar energy collectors, miniature fuel cells, condensers, evaporators, and reactors \cite{japar20, municchi22, sidik17}. Furthermore, the heat removal capacity of flow boiling microchannels is enhanced by using the latent heat of the working fluid due to phase change \cite{Jain2020BubbleGrowth}. Hence, substantial studies have been conducted on flow boiling in microchannels, emphasizing flow pattern transition, bubble dynamics, pressure drop, and heat transfer mechanism \cite{chen20, Fayyadh2017FlowBoiling, Garimella2006ThermalManagement, karayiannis17, Liu2017BubbleTrainMicrochannel, Magnini2017SlugFlowBoiling, Magnini2016FlowBoiling, prajapati17}.

Flow boiling microchannel heat sinks are typically arranged as micro-evaporators, made of conductive material with many parallel microchannels. Conjugate heat transfer (CHT) occurs when the working fluid flows over the solid body conducting heat. Hence, the heat transfer is via conduction within the solid and convection at the solid-fluid interface \cite{municchi22}. As such, the solid-fluid interface temperature and heat flux cannot be predicted in advance because they are dependent on instantaneous fluid flow and solid wall properties. After nucleation begins within microchannels during flow boiling, a growing bubble expands quickly to occupy the whole cross-section, leading to slug flow regimes. Due to its effectiveness, slug flow is considered as an optimal operating condition. In addition, the local heat transfer is significantly enhanced as the liquid film beneath the slug bubble evaporates \cite{jacobi02}.

There have been many studies on flow boiling in microchannels \cite{baldassari13, Bertsch2008ReviewFlowBoiling, garimella03, kadam21, nahar21, thome04}. Many experiments were performed to explore the microchannel flow boiling mechanisms \cite{Bigham2015FlowBoiling, Gedupudi2011BubbleGrowth, halon22, Huh2007FlowBoiling, kumar22, rui23, tang22, Wang2014MicrochannelFlowBoiling, zhang20, zhang17, zhou21}. Numerical simulation, in addition to experiments, serves as a vital tool for comprehending the boiling in microchannels as it can offer many details that cannot be obtained experimentally. Magnini and Thome \cite{Magnini2016FlowBoiling} studied numerically the slug flow regime, and observed that confined bubbles arise when the bubble departure diameter is greater than the channel hydraulic diameter. Lorenzini and Joshi \cite{lorenzini18} conducted numerical simulations and investigated the reduced saturation pressure, sub-cooling, and flow configuration. Ferrari et al. \cite{Ferrari2018SlugFlowBoiling} studied the influence of the cross-sectional geometry of the channel, and found that bubbles travel faster in a square channel than in a circular tube. Luo et al. \cite{luo20} studied the annular flow boiling, and found the thickness of the liquid film between the interface and the heating wall decreases as the wall heat flux or inlet vapor quality increases. Magnini and Matar \cite{magnini20} considered the influence of the aspect ratio of the microchannel, and observed that at low capillary numbers, square microchannels have very thin liquid films, while rectangular microchannels have thicker liquid films as the aspect ratio increases. Siddique et al. \cite{siddique20} developed a model to numerically study the ratio of contact diameter to bubble diameter, and found that this ratio is time-varying. Tian et al. \cite{tian23} considered microchannel flow boiling under rolling conditions and found that bubbles are difficult to coalesce and agglomerate which hinders the formation of regular flow regimes observed under static conditions. Odumosu et al. \cite{odumosu23} compared the bubble growth in straight and wavy channels, and found that the wavy channel can remarkably enhance convection by perturbing the flow field. Guo et al. \cite{guo23} studied numerically the microchannel surface roughness, and found that thermal-hydraulic performance is improved with the presence of surface roughness than without it. Zhang et al. \cite{zhang23} investigated numerically the flow regime transition in vertical microchannel, and developed the phase diagram to predict the boiling stages. The local front reconstruction method was adopted by Rajkotwala et al. \cite{rajkotwala22} to simulate microchannel flow boiling of a vapor bubble, and found that the method can accurately track the bubble interface.

Despite numerous studies on flow boiling in microchannels, previous studies mainly consider the flow of the fluid only, and often neglect the effects of the heat conduction in the solid wall. However, conjugate heat transfer becomes notably significant in microchannel heat sinks due to the nature of heat transfer within solid walls, where it occurs in all directions. Moreover, the comparable sizes of the wall thickness and the channel dimensions further emphasize the importance of considering conjugate heat transfer in such systems. Tiwari and Moharana \cite{tiwari19} investigated conjugate heat transfer of flow boiling in a microtube, and found that a maximum bubble length exists with an optimum conductivity of the channel wall. Lin et al. \cite{lin21} considered the flow boiling in a rectangular microchannel, modeling one evaporator wall for the computational domain. They found that highly conductive materials and thicker walls showed higher temperatures with increased bubble growth rates, enhancing two-phase heat transfer. More recently, Municchi et al. \cite{municchi22} investigated the bubble dynamics, heat transfer, and evaporator temperature in a microchannel, and found that the heat transfer performance of the heat sink is remarkably affected by the conjugate effect. Vontas et al. \cite{vontas22} considered the influence of properties of the solid material, and observed that the two-phase flow patterns and the heat transfer are non-uniform for the same applied heat flux and configuration. Because of the complexity of microchannel flow boiling, it is needed to further explore more details on the bubble dynamics of the conjugate heat transfer in flow boiling microchannels. Therefore, we study the effects of the conjugate heat transfer on flow boiling heat transfer in microchannels, focusing on the dynamics of the bubbles. We consider different solid materials with different solid wall thicknesses and analyze their influence on bubble growth and heat transfer performance.

\section{Numerical method}\label{sec:2}
\subsection{Numerical model}\label{sec:2.1}
We utilize OpenFOAM to simulate the flow boiling with conjugate heat transfer in microchannels. The multiRegionPhaseChangeFlow solver by Scheufler and Roenby \cite{scheufler21} is adopted for the simulation. The solver takes into account the influence of phase change and conjugate heat transfer by solving the mass, momentum, energy, and phase fraction equations. The numerical model is built for both the fluid and solid regions of the domain. The equations of mass, momentum, and energy conservation are solved in the fluid region for the flow of liquid and vapor. In contrast, only the energy equation is solved within the solid region. The mass and momentum conservation equations for the fluid region are
\begin{equation}\label{eq:01}
  \frac{\partial \rho}{\partial t}+\nabla \cdot (\rho \mathbf{u})=0
\end{equation}
\begin{equation}\label{eq:02}
  \frac{\partial (\rho \mathbf{u})}{\partial t}+\nabla \cdot (\rho \mathbf{uu})=-\nabla {{p}}+\nabla \cdot \{{{\mu }_{\text{eff}}}[\nabla \mathbf{u}+{{(\nabla \mathbf{u})}^\textbf{T}}]\}+\rho \textbf{g}+{\mathbf{f}}_{\sigma }
\end{equation}
where ${{\mu }_{\text{eff}}}$ is the fluid viscosity, and ${{\mathbf{f}}_{\sigma }}$ is the surface tension force.
The energy equation adopts a two-field approach in terms of the temperature, ${{T}}$
\begin{equation}\label{eq:04}
  \frac{\partial ({{\rho }}c_{p}{{T}})}{\partial t}+\nabla \cdot ({{\rho }}c_{p}\mathbf{u}{{T}})=\nabla \cdot ({{k}}\nabla {{T}})+{{\dot{q}}}_{pc}
\end{equation}
where $k$ is the thermal conductivity, ${{c}_{p}}$ is the specific heat capacity, and ${{\dot{q}}}_{pc}$ is for energy changes caused by phase change. For the solid region, Eq.\ (\ref{eq:04}) becomes the heat conduction equation in Eq.\ (\ref{eq:10}).

The volume of fluid (VOF) method is adopted to predict the evolution of the liquid-vapor interface,
\begin{equation}\label{eq:05}
  \frac{\partial \alpha }{\partial t}+\nabla \cdot (\mathbf{u}\alpha )={{\dot{\alpha }}_{pc}}
\end{equation}
where $\alpha$ is the volume fraction of the liquid phase, and ${{\dot{\alpha }}_{pc}}$ is the explicit source term to account for phase change.

The thermophysical parameters of the fluid are calculated from the volume fraction
\begin{equation}\label{eq:06}
  \varphi =(1-\alpha ){{\varphi }^{v}}+\alpha {{\varphi }^{l}}
\end{equation}
where $\varphi $ represents any fluid properties, including viscosity, density, and thermal conductivity.

The surface tension effect is modeled by including a source term ${{\mathbf{f}}_{\sigma }}$ in Eq.\ (\ref{eq:02}) at the liquid-vapor interface \cite{Brackbill1992ModelingSurfaceTension}
\begin{equation}\label{eq:07}
  {{\mathbf{f}}_{\sigma }}=\sigma \kappa \mathbf{n}\left| \nabla \alpha  \right|\frac{2\rho }{{{\rho }^{v}}+{{\rho }^{l}}}
\end{equation}
where $\kappa $ and $\mathbf{n}$ are the curvature and the unit normal vector of the liquid-vapor interface.
\begin{equation}\label{eq:08}
  \kappa =-\nabla \cdot \mathbf{n}
\end{equation}
\begin{equation}\label{eq:09}
  \mathbf{n}=\frac{\nabla \alpha }{\left| \nabla \alpha  \right|}
\end{equation}

The mass transfer due to phase change at the liquid-vapor interface was caculated using the Hardt and Wandra model \cite{hardt08} with the VOF equation to account for the liquid loss. The model was developed to simulate microscale phase change processes and can improve the numerical stability. For more details of the phase-change model, the readers can refer to Refs.\ \cite{hardt08, scheufler21}.

The energy conservation equations for the solid and fluid regions are solved separately on different meshes and connected at the solid-fluid interface via a common boundary. At the solid-liquid interface, the temperature and the heat flux on the solid side are equal to that on the fluid side. The energy equation for the solid region is simply a heat conduction equation:
\begin{equation}\label{eq:10}
  \frac{\partial ({{\rho }^{s}}{{c_p^s}T^{s}})}{\partial t}=\nabla \cdot ({{{k }^{s}}}\nabla {T^{s}})
\end{equation}
where ${{\rho }^{s}}$ and $c_{p}^{s}$ are the density and specific heat capacity of the solid material of the wall of the microchannels, respectively.

\subsection{Simulation setup}\label{sec:2.2}
\begin{figure}
  \centering
  \includegraphics[scale=0.6]{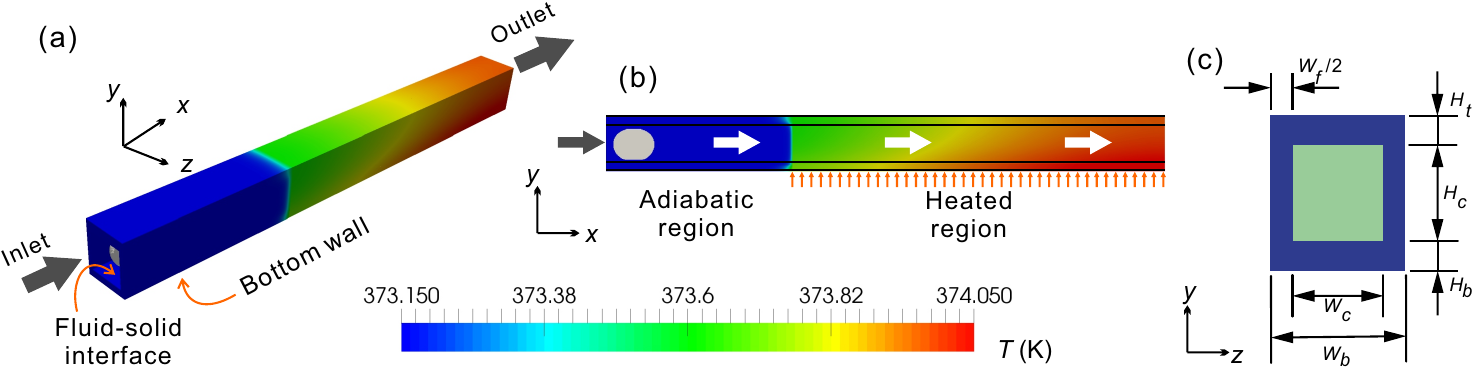}
  \caption{Simulation setup for the conjugate heat transfer of flow boiling in a microchannel. (a) Full geometry of the channel. (b) Sectional view to show the bubble and the internal temperature distribution in the middle cross-section (perpendicular to the $z$-direction) with white arrows showing the continuous phase moving in the channel. (c) Dimensions of the cross-section perpendicular to the $x$-direction.}\label{fig:01}
\end{figure}

Multi-microchannel heat sinks include many microchannels in parallel fabricated onto a block of a thermal conductive material, and are typically heated from below \cite{municchi22}. Usually, a surface of the heat sink is heated by a heat source that exerts a specific amount of heat generation. This numerical work considers a microchannel unit including a channel and four solid walls on the four sides imitating the geometry of a heat sink with multiple microchannels \cite{szczukiewicz13}, as illustrated in Figure \ref{fig:01}.

The microchannel has a square cross-section of 200 $\times$ 200 $\upmu$m$^2$ and a channel length of 3 mm. Hence, it has a hydraulic diameter of $D_h = 200$ $\upmu$m. The thickness and properties of the wall are varied to analyze their effect. The microchannel is divided into two sections, an adiabatic section of $L_a = 5D_h$ for the development of the bubble in the microchannel to achieve a steady movement, and a heated section of $L_h = 10D_h$ where a constant heat flux is implemented at the bottom base solid. The adiabatic section allows the bubble to achieve a stabilized position and shape in the microchannel before entering the heat section. The bottom wall thicknesses $H_b$ are varied as 20, 40, 80, and 160 $\upmu$m. The top wall thickness $H_t$ and vertical sidewall thickness $W_f$ are constant with $D_h/4$. Symmetry boundary conditions are implemented at the half of the side wall thickness to mimic the existence of adjacent microchannels in parallel within a microchannel heat sink. To reduce the computational demand, only half of the microchannel unit is simulated by using a symmetry boundary condition at the middle cross-section of the $z$-direction at $z = 0$.

Four solid domain materials, silicon (Si), aluminum (Al), copper (Cu), and magnesium (Mg) are considered with their properties listed in Table \ref{tab:01}. Single-phase simulations with only liquid are performed first to achieve the fully developed steady-state, and the velocity and temperature fields are taken as initial conditions for the multiphase simulations. Then at $t$ = 0, a vapor bubble is initiated as a cylinder with $1.1D_h$ length, $0.8D_h$ axial diameter and spherical caps, as shown in Figure \ref{fig:01}. At the inlet of the microchannel, the liquid has a velocity of 0.294 m/s and a saturation temperature $T_\text{sat}$. A uniform heat flux $q_w = 30$ kW/m$^2$ is implemented at the heated region of the bottom base wall only and not including the sidewalls of the microchannels. This is to imitate the existence of adjacent microchannels in parallel within a multi-microchannel heat sink with a heat source on the bottom base. The properties of the working fluid (i.e., water) are listed in Table \ref{tab:02}. Due to a minimal temperature change near the saturation condition, constant surface tension is considered in this study because the impact of the temperature variation on the surface tension is negligible.

\begin{table*}[]
\footnotesize
\center
\caption{Properties of channel materials used in this study.}
\label{tab:01}
\begin{tabular}{lllll}
\hline
Property                                             & Silicon & Aluminum & Copper & Magnesium \\ \hline
Density, $\rho$ [kg/m$^3$]                       & 2300    & 2700     & 8940   & 1740      \\
Specific heat capacity, $c_p$   [J/(kg$\cdot$K)] & 700     & 880      & 385    & 1004      \\
Thermal conductivity, $k$ [W/(m$\cdot$K)]       & 150     & 237      & 398    & 156       \\
Thermal diffusivity, $\alpha_T$ [mm$^2$/s]          & 93.17   & 99.75    & 116    & 86.2      \\ \hline
\end{tabular}
\end{table*}

\begin{table}[]
\footnotesize
\center
\caption{Properties of working fluid (i.e., water) used in this study.}
\label{tab:02}
\begin{tabular}{lll}
\hline
Property                                             & Liquid                & Vapor                  \\ \hline
Density, $\rho$ [kg/m$^3$]                          & 958.4                 & 0.598                  \\
Dynamic viscosity, $\mu$ [Pa$\cdot$s]            & $2.817 \times10^{-4}$ & $1.227 \times 10^{-5}$ \\
Specific heat capacity, $c_p$   [J/(kg$\cdot$K)] & 4216                  & 419                    \\
Thermal conductivity, $k$ [W/(m$\cdot$K)]       & 0.67909               & 0.025096               \\
Saturation temperature, $T_\text{sat}$   [K]     & 373.15                &                        \\
Surface tension, $\sigma$ [N/m]                  & 0.0588                &                        \\
Latent heat, $h_{lv}$ [kJ/kg]                    & 2230                  &                        \\ \hline
\end{tabular}
\end{table}

The average Nusselt number is used to characterize the convection at the solid-fluid interface of the heated bottom wall ${{\overline{\Nu}}_{\text{bottom}}}$, and it is calculated from the average convective heat transfer coefficient ${{\overline{h}}_{\text{bottom}}}$,
\begin{equation}\label{eq:11}
  {{\overline{\Nu}}_{\text{bottom}}}=\frac{{{{\bar{h}}}_{\text{bottom}}}{{D}_{h}}}{{{k}^{l}}}
\end{equation}
\begin{equation}\label{eq:12}
  {{\bar{h}}_{\text{bottom}}}=\frac{1}{A}\int_{0}^{A}{{{h}_{\text{bottom}}}dA}
\end{equation}
\begin{equation}\label{eq:13}
  {{h}_{\text{bottom}}}=\frac{{{q}_{\operatorname{int}}}}{{{T}_{\mathrm{int}}}-{{T}_{\text{sat}}}}
\end{equation}
where ${{q}_{\operatorname{int}}}$ and $T_\text{int}$ are the local heat flux and local temperature at the solid-fluid interface of the heated bottom wall.

\subsection{Mesh independence study}\label{sec:2.3}
\begin{figure}
  \centering
  \includegraphics[scale=0.55]{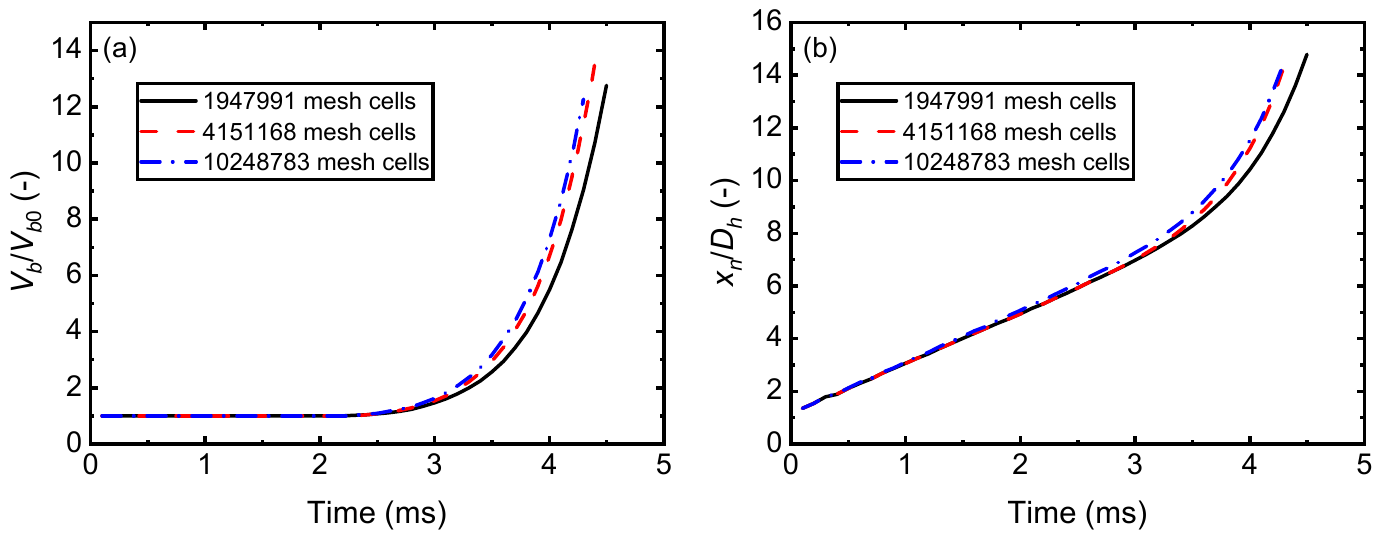}
  \caption{Mesh independence study: (a) dimensionless bubble volume $V_b/V_{b0}$, (b) dimensionless position of the bubble nose. The bottom wall is silicon with a thickness of $H_b = 40$ $\upmu$m.}\label{fig:02}
\end{figure}

To resolve the flow details in the microchannel, finer mesh is used in the fluid region than the solid region with ${{\Delta }_{\text{fluid}}}={{{\Delta }_{\text{solid}}}}/{2}$, where ${{\Delta }_{\text{fluid}}}$ and ${{\Delta }_{\text{solid}}}$ are the grid sizes in the fluid and solid regions, respectively. To balance the simulation cost and accuracy, a mesh independence study is performed. Three different mesh schemes of 1947991, 4151168, and 10248783 cells are tested for Silicon material with 40 $\upmu$m bottom wall thickness. The bubble volume and nose position are used in the evaluation of the grid independency, which are two parameters to quantify the bubble growth dynamics. As shown in Figure \ref{fig:02}, there is only a minute difference between the results of the mesh schemes of 4151168 and 10248783 cells. Therefore, the mesh with 4151168 cells is adopted for the subsequent simulations.

\subsection{Validation}\label{sec:2.4}
\begin{figure}
  \centering
  \includegraphics[scale=0.32]{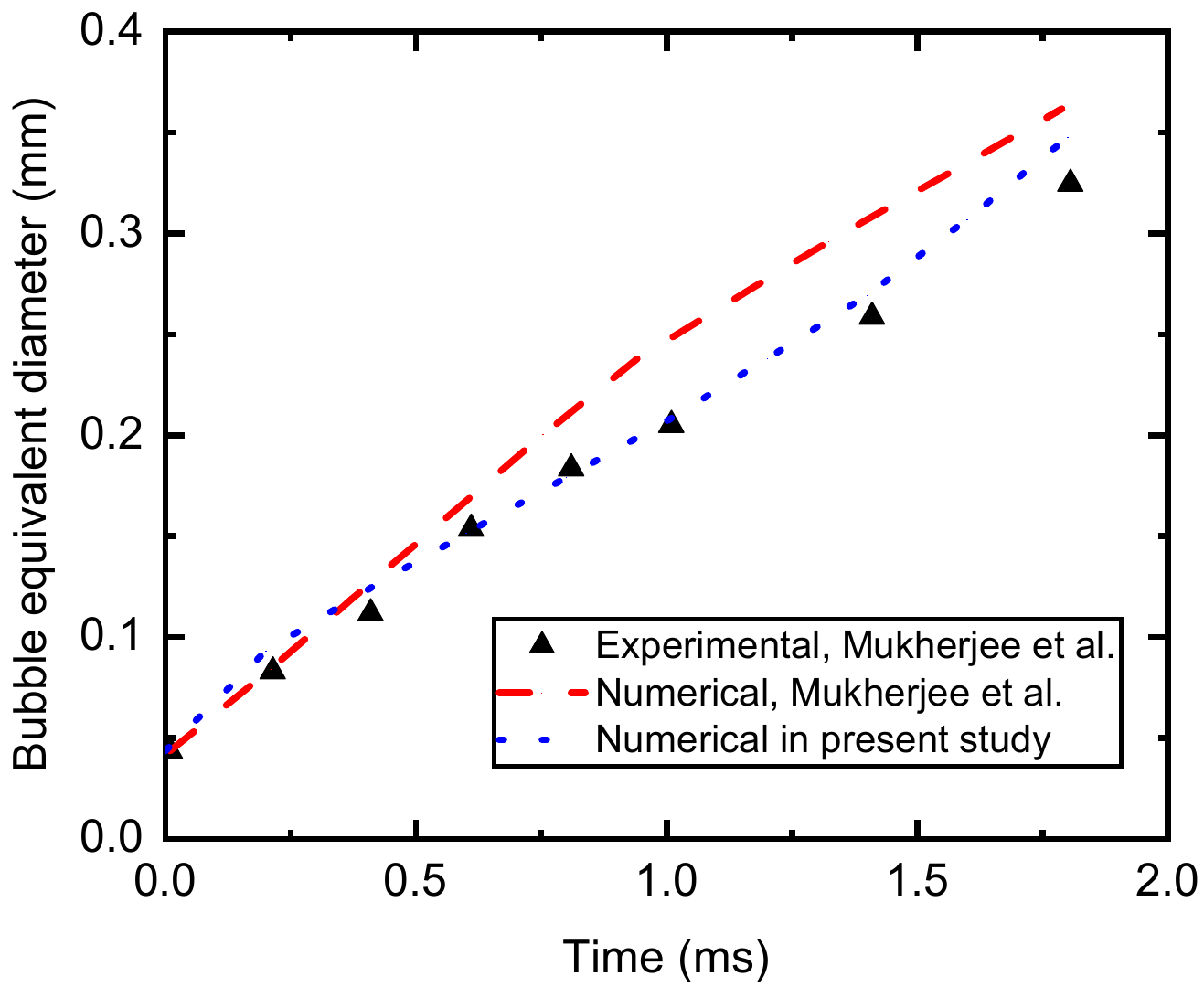}
  \caption{Validation of the model by comparing the bubble equivalent diameter, $d_{eq} = \left( {6V_b}/{\pi} \right)^{{1}/{3}}$ with experimental results in Ref.\ [\citenum{mukherjee11}].}\label{fig:03}
\end{figure}

The simulation is validated against the experimental results of flowing bubble growth in a microchannel by Mukherjee et al.\ \cite{mukherjee11}. As shown in Figure \ref{fig:03}, the equivalent bubble diameter computed in the current study agrees well with the experimental data, which proves the validity and accuracy of our model.

\section{Results and discussion}\label{sec:3}
\subsection{Typical process of conjugate heat transfer}\label{sec:3.1}
\begin{figure}
  \centering
  \includegraphics[scale=0.5]{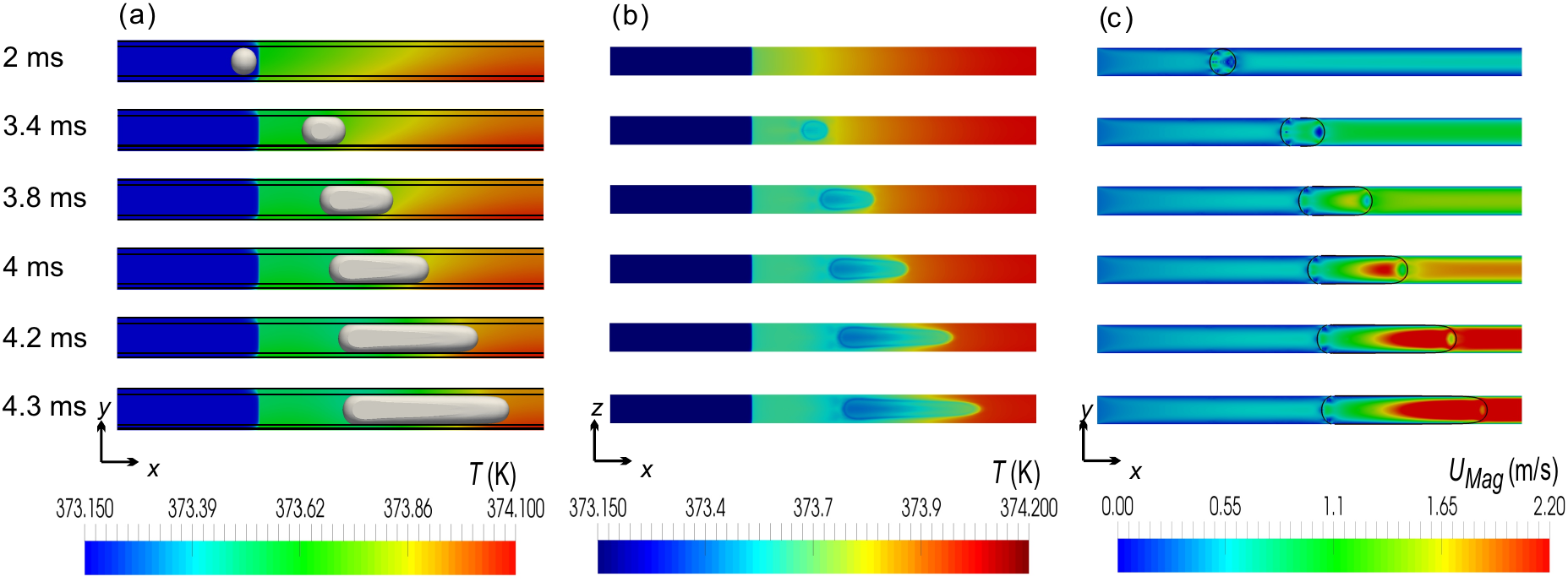}
  \caption{Time evolution of the flow boiling conjugate heat transfer in a typical microchannel: (a) bubble shapes and temperature fields in the middle cross-section in the $z$-direction; (b) temperature fields at the solid-fluid interface of the bottom wall; (c) velocity field at the middle cross-section in the $z$-direction. The bottom wall is silicon with a thickness of $H_b = 40$ $\upmu$m.}\label{fig:04}
\end{figure}

	A typical process of flow boiling conjugate heat transfer of a single bubble in a microchannel is shown in Figure \ref{fig:04}. The bubble is transported downstream by the liquid as the liquid flows in the microchannel. When the bubble is in the adiabatic region, the bubble size remains constant because there is no heat transfer from the solid to the fluid and no vaporization from liquid to vapor, as shown in Figure \ref{fig:04}(a). When the bubble reaches the heated region, the bubble gains heat from the heated liquid and begins to grow. Particularly, when the bubble contacts the superheated thermal boundary layer, the bubble grows quickly as the liquid evaporates. Hence the bubble nose elongates remarkably and accelerates downstream. Given the confinement effect of the wall, the bubble mainly elongates in the flow direction, but its expansion in the spanwise direction is negligible. The temperature fields at the solid-fluid interface of the heated bottom wall are shown in Figure \ref{fig:04}(b). Before the bubble reaches the heated region, the temperature at the solid-fluid interface increases along the channel. As the bubble starts to grow, it forms a contact region over the bottom wall, which reduces the local temperature on the wall because of the rapid vaporization. Therefore, as the bubble grows and moves downstream, the contact region becomes larger and the local solid-fluid interface temperature reduces further.

\begin{figure}
  \centering
  \includegraphics[scale=0.35]{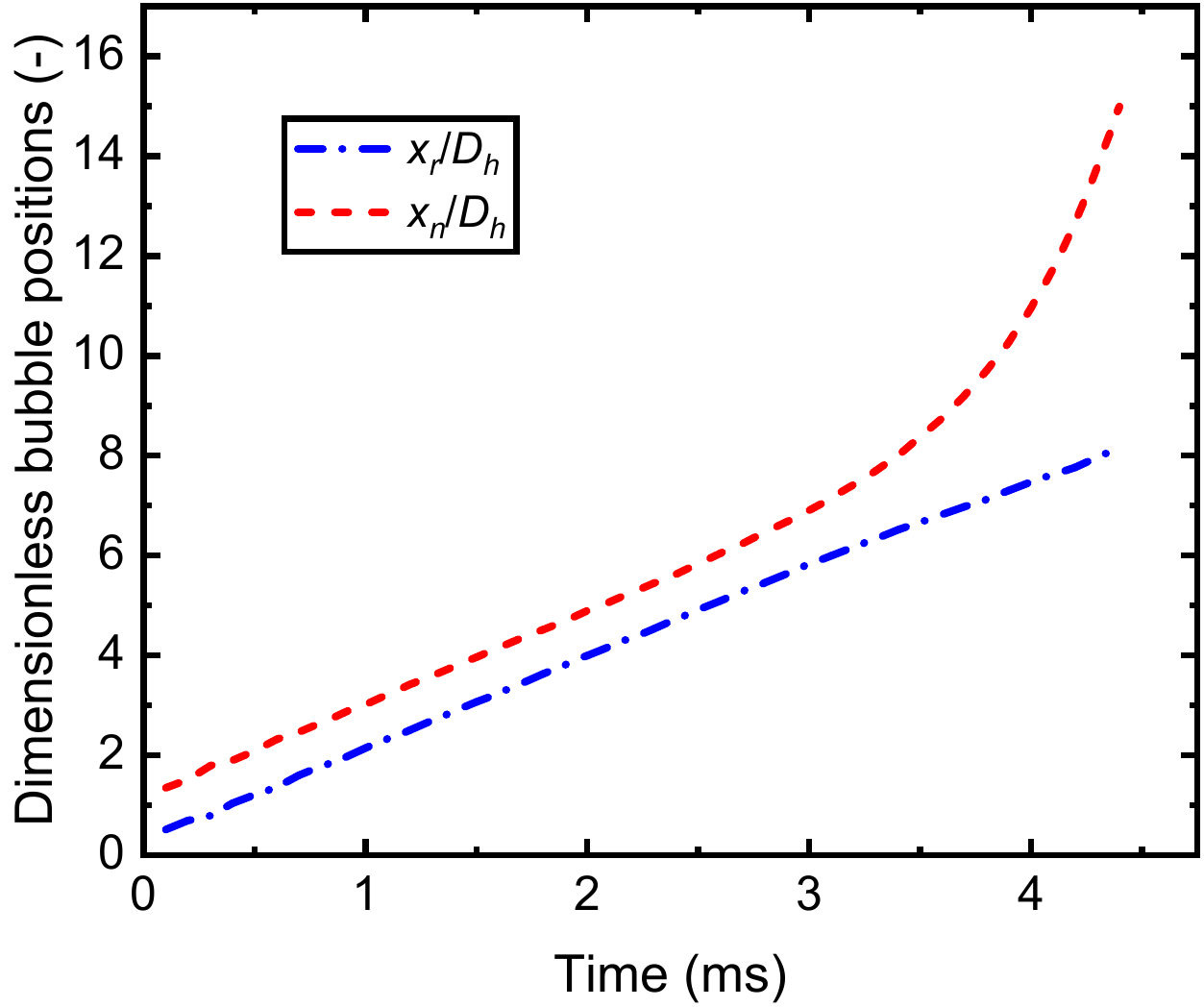}
  \caption{Time variation of the dimensionless bubble positions (the rear $x_r$ and the nose $x_n$).}\label{fig:05}
\end{figure}

The corresponding velocity fields are shown in Figure \ref{fig:04}(c). We can see that the flow in the microchannel is remarkably perturbed by the bubble. When the bubble is in the adiabatic section, only the velocity in the very small area near the bubble is affected. The bubble produces perturbation to the flow field, which enhances the thermal convection from the wall to the fluid as the bubble enters the heated region \cite{che20122d,che20153d}. In the heated region, the bubble quickly expands, pushing the liquid in front of the bubble forward. Therefore, we can see that the fluid velocity in the front part of the bubble is much higher than the velocity in the rear part. The liquid in front of the bubble is also pushed by the bubble, hence it has a velocity larger than that behind the bubble. As the bubble size further grows, the difference in the velocity between the front and rear parts of the bubble further increases. This can be quantified by the time variation of the nose and the rear of the bubble, as shown in Figure \ref{fig:05}.

\begin{figure}
  \centering
  \includegraphics[scale=0.35]{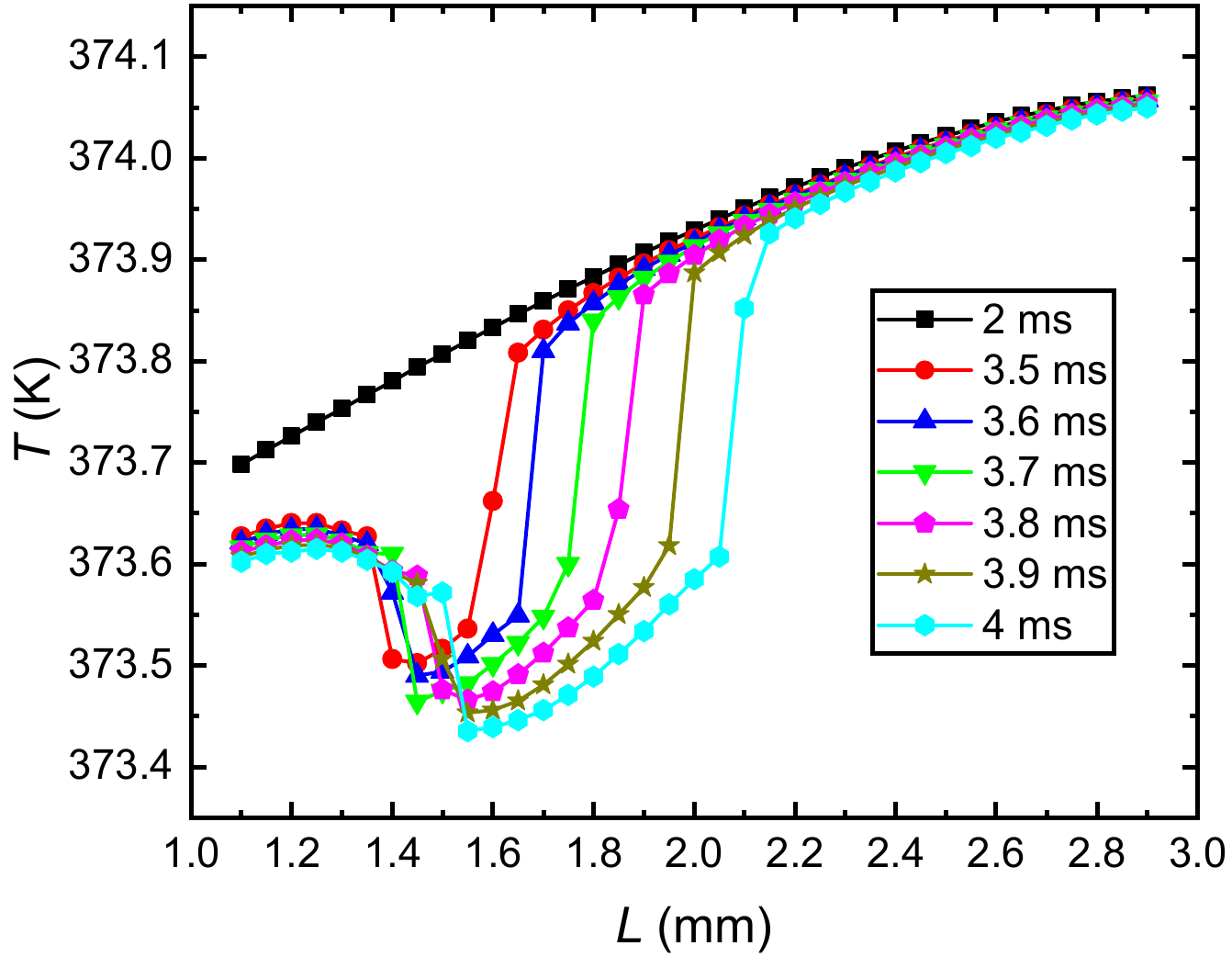}
  \caption{Time variation of the temperature profiles at the solid-fluid interface of the heated bottom wall ($z = 0$). The bottom wall is silicon with a thickness of $H_b = 40$ $\upmu$m.}\label{fig:06}
\end{figure}

To quantitatively evaluate the effect of the bubble presence when the bubble passes through the channel, the time variation of the temperature distribution along the microchannels in the heated region is shown in Figure \ref{fig:06}. Before the arrival of the bubble, the fluid temperature increases monotonically along the axial direction of the microchannel in the heated region. However, as the bubble arrives, it creates a strong perturbation to the flow field and the temperature field in the liquid, hence it cools the local temperature of the microchannel wall. Therefore, we can see a rapid drop in the wall temperature. As the bubble proceeds, the cooling effect of the bubble continues, further reducing the wall temperature. Therefore, the solid temperature at the heated region declines as the bubble expands and passes through. This result indicates a strong coupling between fluid and solid, highlighting the necessity of taking account of the heat conduction in the solid wall when analyzing the flowing boiling process.

\subsection{Effect of bottom wall thickness}\label{sec:3.2}
\begin{figure}
  \centering
  \includegraphics[scale=0.5]{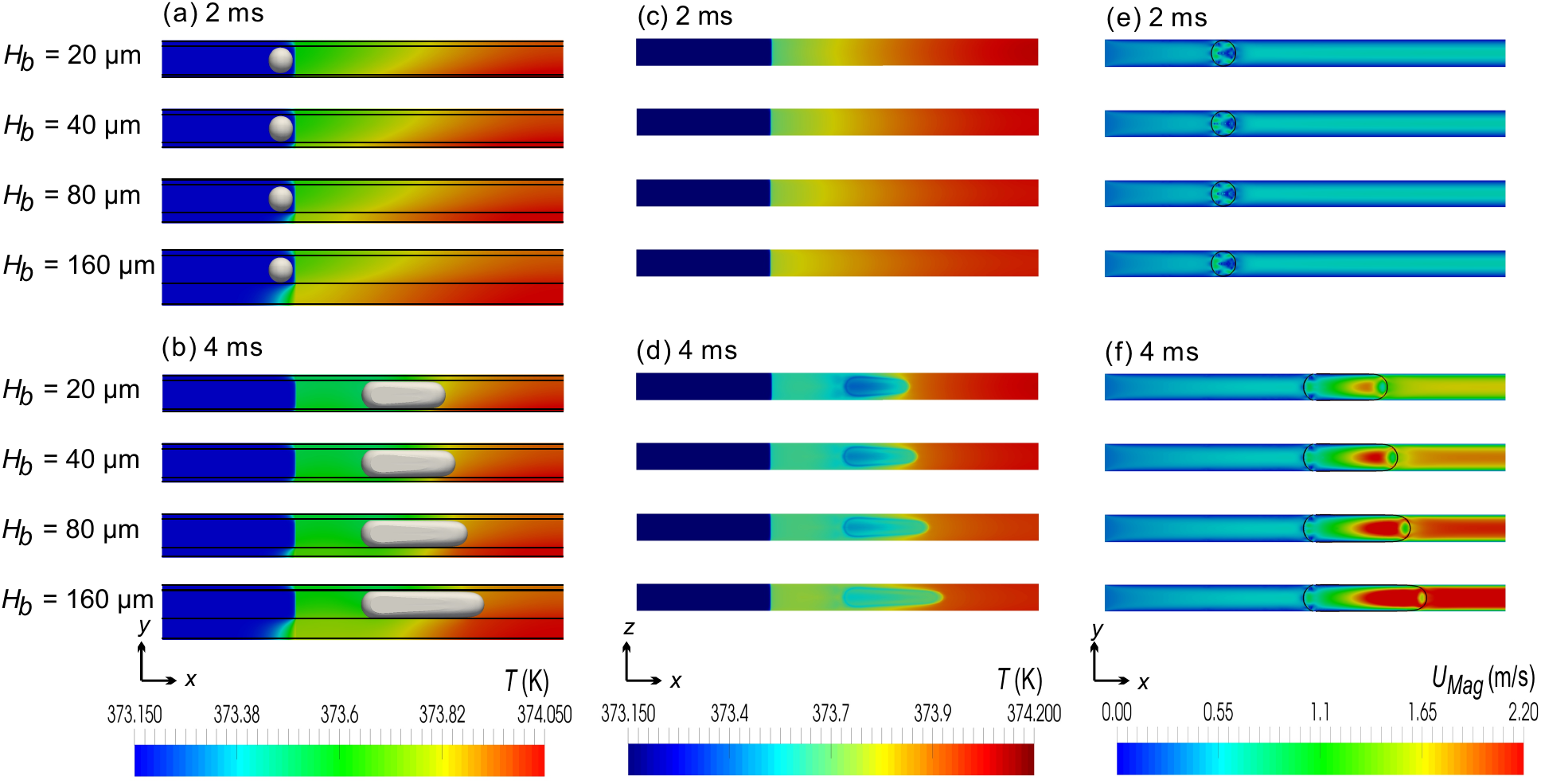}
  \caption{Flow boiling in silicon microchannels with different bottom wall thicknesses: (a, b) bubble shapes and temperature fields in the middle cross-section in the $z$-direction; (c, d) temperature fields at the solid-fluid interface of the bottom wall; (e, f) velocity field at middle cross-section in the $z$-direction. Panels (a, c, e) show results when the bubbles are in the adiabatic region ($t$ = 2 ms), while panels (b, d, f) show results when the bubbles are in the heated region ($t$ = 4 ms).}\label{fig:07}
\end{figure}

To consider the conjugate heat transfer effect of different bottom wall thicknesses on the boiling heat transfer performance in microchannels, Figure \ref{fig:07} displays the expansion of a bubble and the temperature distribution of the internal solid wall at different bottom wall thicknesses. In the adiabatic region, the bubble size keeps constant for all channels, as shown in Figure \ref{fig:07}(a). The influence of different bottom wall thicknesses on the bubble expansion becomes pronounced in the heat region, see Figure \ref{fig:07}(b). This suggests that the bubble expansion is non-uniform for different bottom wall thicknesses even the wall heat flux is the same. Surprisingly, as the bottom wall becomes thicker, the bubbles grow faster. This result is somewhat counter-intuitive, because one may think that a thicker wall leads to a larger thermal resistance of conduction, hence reducing the heat flux to the liquid and slowing down the bubble growth. Indeed, this thought is not true because of the constant-heat-flux boundary condition at the bottom wall.

\begin{figure}
  \centering
  \includegraphics[scale=0.6]{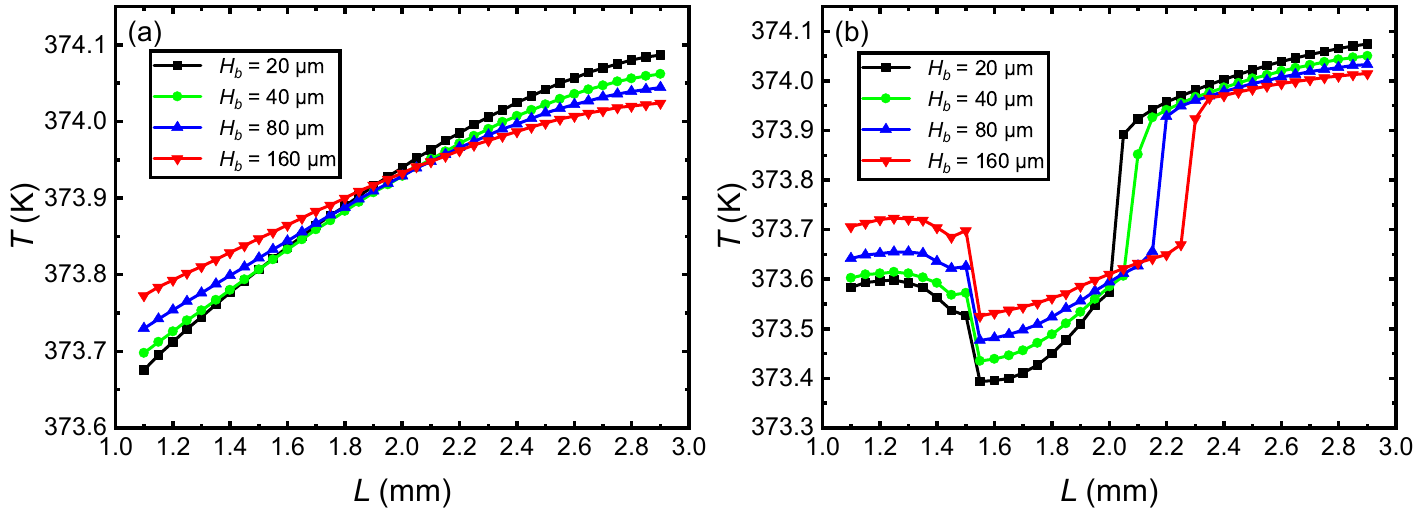}
  \caption{Temperature distribution for different bottom wall thicknesses at the solid-fluid interface of the heated bottom wall ($z$ = 0). (a) Temperature distribution when the bubbles are in the adiabatic region ($t$ = 2 ms), (b) Temperature distribution when the bubbles are in the heated region ($t$ = 4 ms). The bottom wall is silicon.}\label{fig:08}
\end{figure}

To elucidate the underlying mechanism of the faster bubble growth at thicker bottom walls, we plot the temperature distribution at the solid-fluid interface of the heated bottom wall for different bottom wall thicknesses, see Figure \ref{fig:08}. The temperature profiles along the centerline of the solid-fluid interface of the heated bottom wall are presented at two typical instants. The instant of $t$ = 2 ms is before the bubble arrives at the heated region, while the instant of $t$ = 4 ms is when the bubble is in the heated region and has produced a cooling effect on the channel. By comparing the temperature profiles at $t$ = 2 ms, we can see that, even though the temperature all increases monotonically, the rates of the temperature increase along the flow direction are different and affected by the bottom wall thickness. A thicker bottom wall results in a slower increase rate and a higher temperature in the upstream. This is mainly because the wall has a higher thermal conductivity than the fluid, hence the bottom wall can spread heat effectively along the flow direction. Therefore, for microchannels with thicker walls, the temperature distribution is more uniform along the channel. As a consequence, the upstream temperature is higher for microchannels with thicker bottom wall. When the vapor bubble arrives at the heated region, the higher temperature leads to a higher degree of superheating, hence increasing the rate of fluid vaporization and also the growth rate of the bubble size.

\begin{figure}
  \centering
  \includegraphics[scale=0.35]{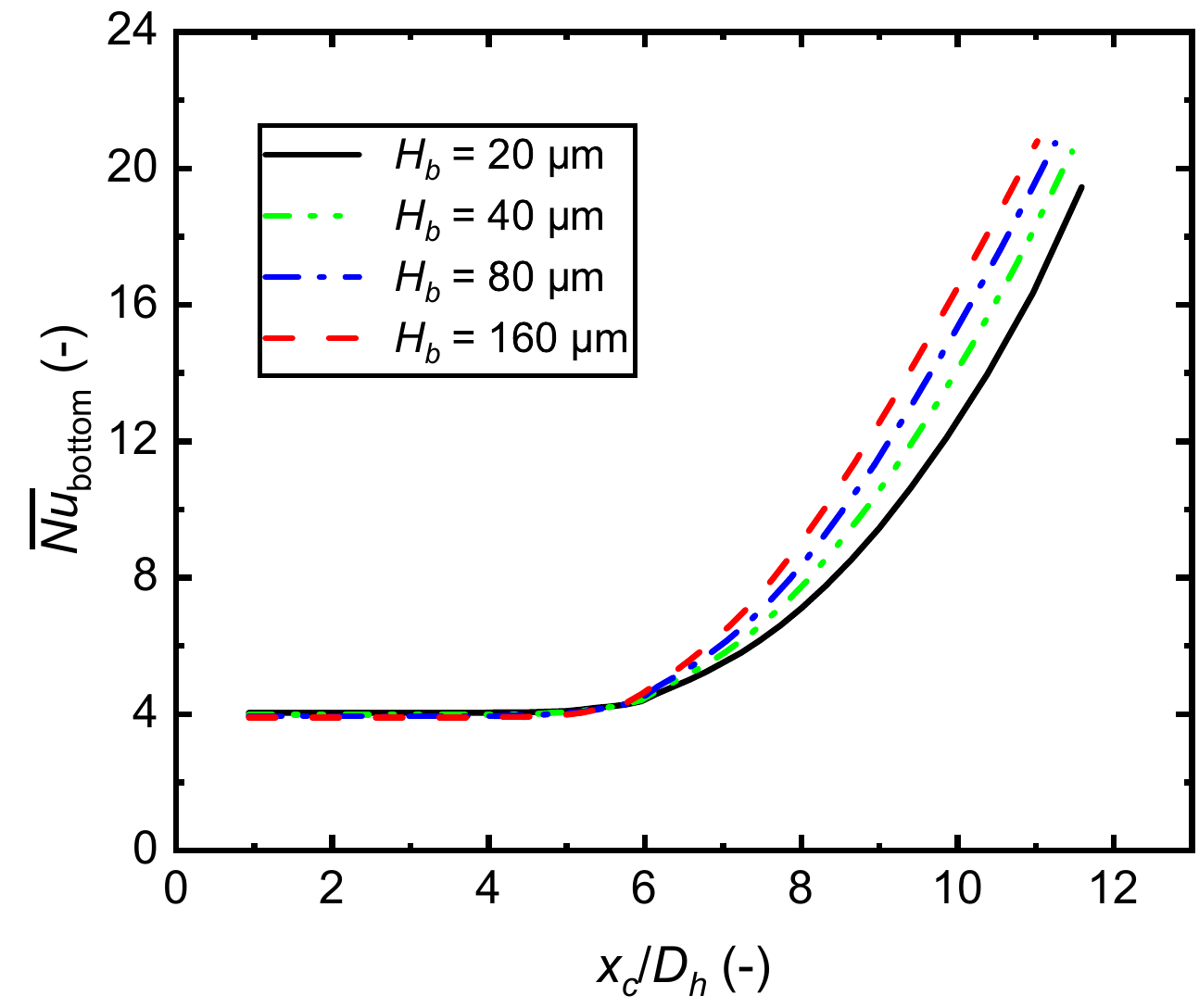}
  \caption{Nusselt number ${{\overline{\Nu}}_{\text{bottom}}}$ of a silicon microchannel plotted versus the dimensionless bubble position $x_c/D_h$ for different bottom wall thicknesses.}\label{fig:09}
\end{figure}

As a consequence of the bubble growth, the Nusselt number for microchannels is remarkably affected by the bottom wall thickness. The average Nusselt number at the solid-fluid interface of the heated bottom wall ${{\overline{\Nu}}_{\text{bottom}}}$ is plotted against the dimensionless location of the bubble $x_c/D_h$ in Figure \ref{fig:09}. The Nusselt number is small and remains constant initially before the bubble reaches the heated section. Then the Nusselt number quickly increases in the heated section as vaporization begins. Because the microchannels with thicker walls produce larger bubbles, hence the large bubbles significantly perturb the flow in the microchannel. Therefore, the convection effect is also enhanced, and the Nusselt number increases significantly.

\begin{figure}
  \centering
  \includegraphics[scale=0.6]{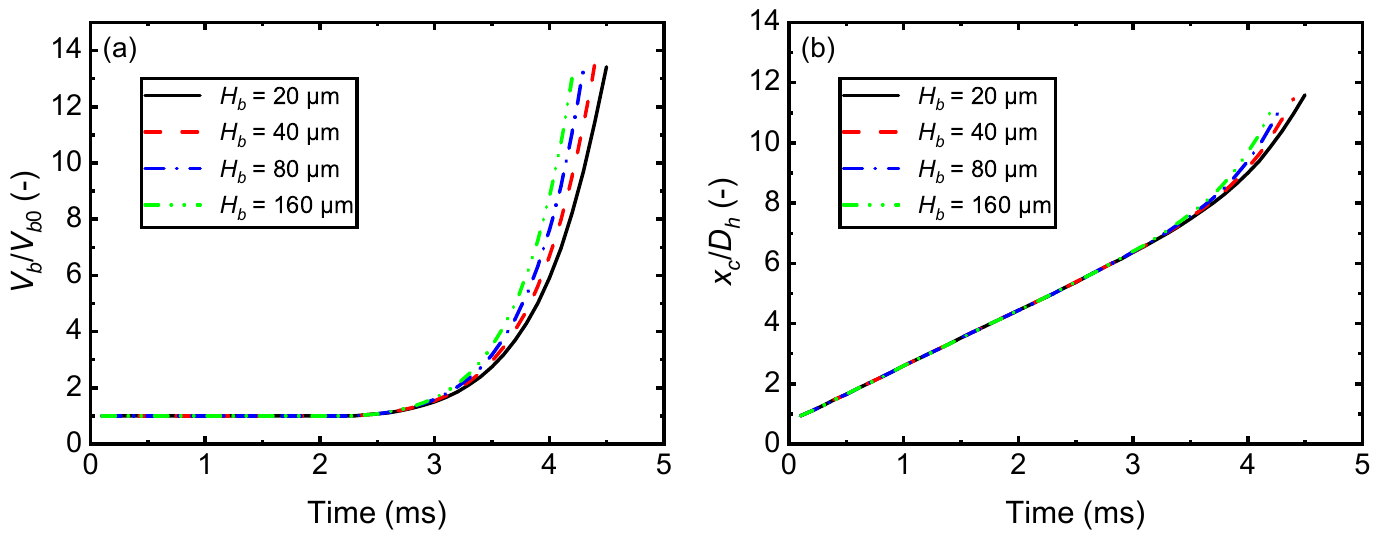}
  \caption{(a) Dimensionless bubble volume $V_b/V_{b0}$ and (b) dimensionless bubble position $x_c/D_h$, plotted versus time for different bottom wall thicknesses.}\label{fig:10}
\end{figure}

To evaluate the effect of the wall thickness on the expansion of the bubble, the dimensionless volume and center position of the bubble at different bottom wall thicknesses are plotted in Figure \ref{fig:10}. A larger bottom thickness results in the fastest bubble growth, see Figure \ref{fig:10}(a). The time variation of the bubble position is presented in Figure \ref{fig:10}(b) by plotting the bubble center position. The bubble center position is calculated by analyzing the position of the bubble centroid from the integrated vapor phase cell volume expressed as 
\begin{equation}\label{eq:14}
x_c = \frac{\int_{V} (1 - \alpha) x \, dV}{\int_{V} (1 - \alpha) \, dV}
\end{equation}
where $x$ is the cell coordinate in the flow direction. At first, the bubble flows at a constant speed in the adiabatic region. Then, the bubble moves faster after reaching the heated region, and the acceleration is influenced by the thickness of the bottom wall. The acceleration is much faster in microchannels with thicker bottom walls than that with thin bottom walls. The acceleration is caused by the bubble expansion under vaporization as the bubble is translated forward. For microchannels with thinner bottom walls, the bubble grows at slower rates compared with that of thicker bottom walls.

\subsection{Effect of different solid materials}\label{sec:3.3}
\begin{figure}
  \centering
  \includegraphics[scale=0.5]{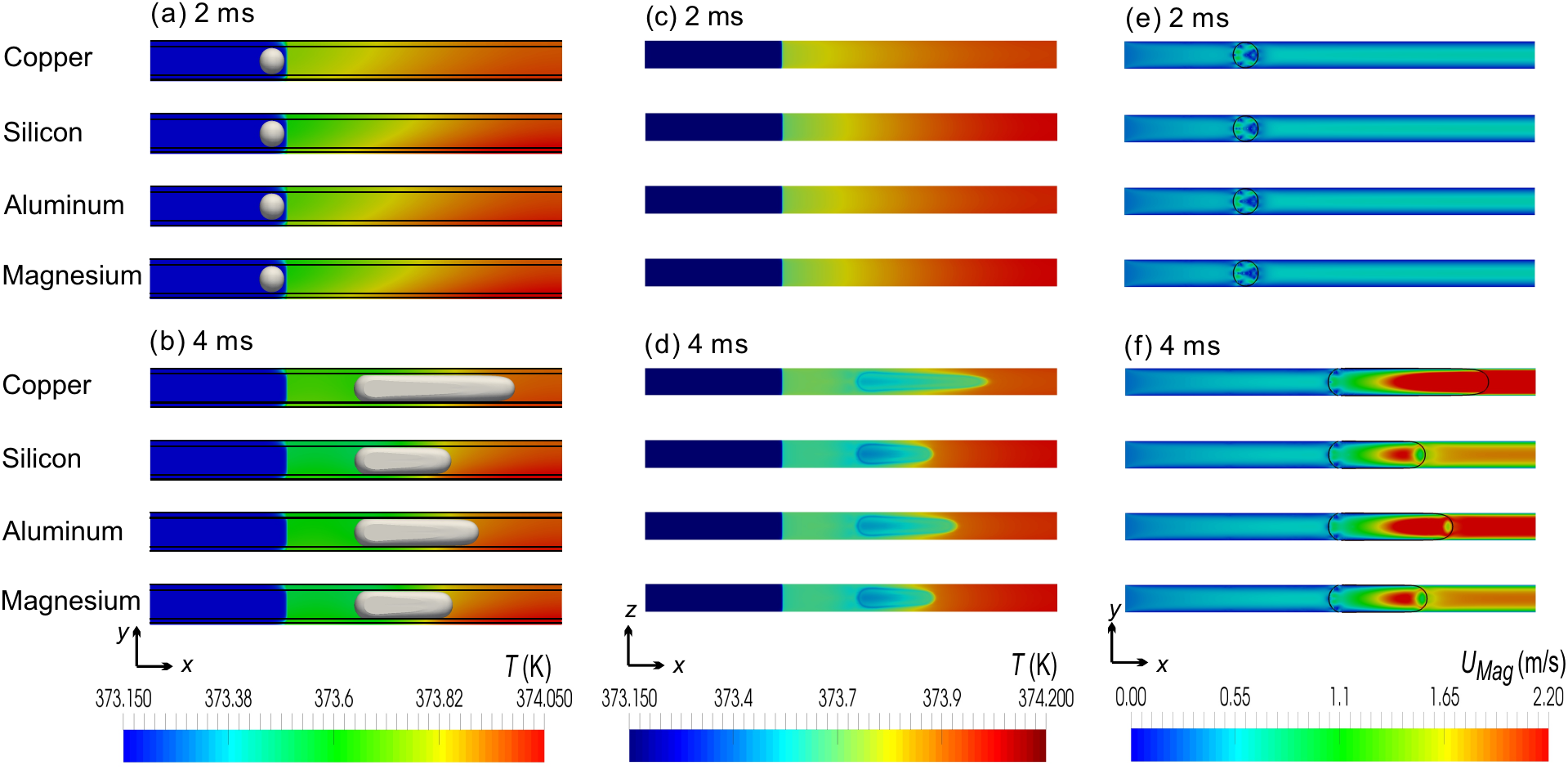}
  \caption{Flow boiling heat transfer in microchannels with different bottom solid materials, e.g., copper, silicon, aluminum, and magnesium: (a, b) bubble shapes and temperature fields in the middle cross-section in the $z$-direction; (c, d) temperature fields at the solid-fluid interface of the bottom wall; (e, f) velocity field at the middle cross-section in the $z$-direction. Panels (a, c, e) show results when the bubbles are in the adiabatic region ($t$ = 2 ms), while panels (b, d, f) show results when the bubbles are in the heated region ($t$ = 4 ms). The bottom wall thickness is $H_b = 40$ $\upmu$m.}\label{fig:11}
\end{figure}

\begin{figure}
  \centering
  \includegraphics[scale=0.6]{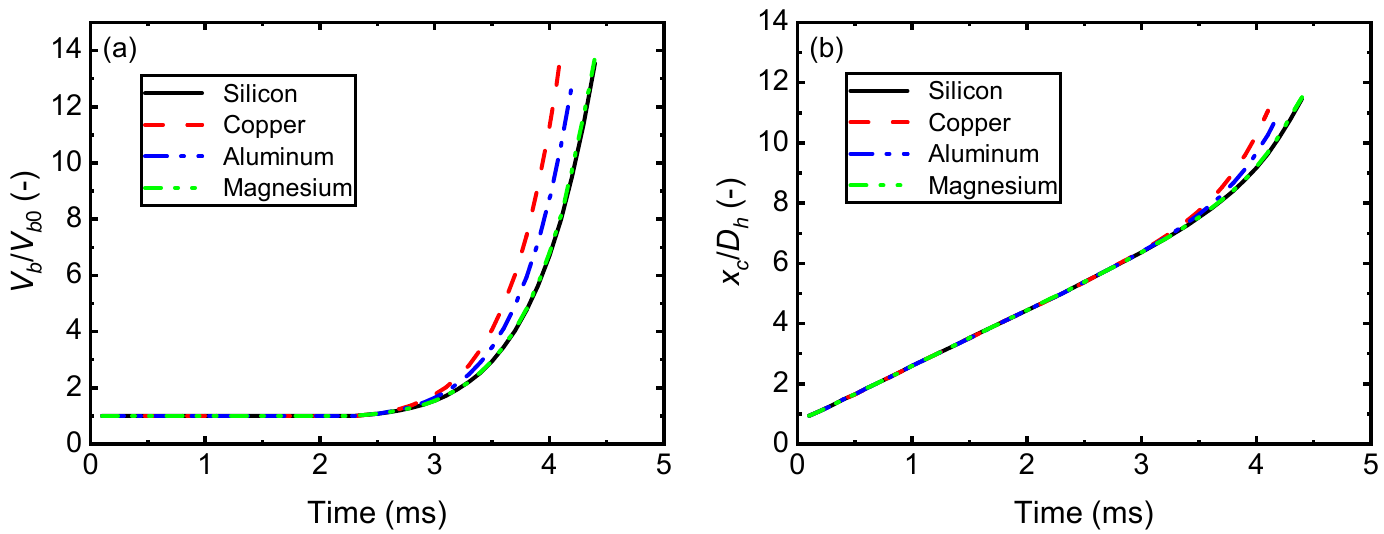}
  \caption{(a) Dimensionless bubble volume $V_b/V_{b0}$ and (b) dimensionless bubble position $x_c/D_h$, plotted versus time for different solid materials. The bottom wall thickness is $H_b = 40$ $\upmu$m.
}\label{fig:12}
\end{figure}

To explore the effects of the solid materials on the conjugate heat transfer, simulations were conducted with several typical solid materials, i.e., copper, silicon, aluminum, and magnesium. The temperature fields and flow fields are presented in Figure \ref{fig:11}. The copper channel has the largest bubble growth even with the same heat flux at the wall, while the silicon and magnesium microchannels have small bubbles. This trend can be observed quantitatively in Figure \ref{fig:12}(a) for the variation of the bubble size. The same trend can also be observed from the curves of the bubble position in Figure \ref{fig:12}(b). The copper microchannel has the fastest bubble acceleration, while the silicon and magnesium microchannels have small bubble accelerations. This consistency is mainly because the bubble quickly grows in the heat region, hence the bubble expansion contributes significantly to the bubble position in the microchannels. The difference in the size and position of the bubble for microchannels with different materials can be attributed to the thermal diffusivity of the materials. The material with a higher thermal diffusivity (for example, copper, as shown in Table \ref{tab:01}) has a higher heat transfer. Hence, copper (with a high thermal diffusivity) can diffuse thermal energy effectively from the heat source to the solid-fluid interface, while magnesium and silicon (with low thermal diffusivities) diffuse thermal energy much more slowly.

\begin{figure}
  \centering
  \includegraphics[scale=0.32]{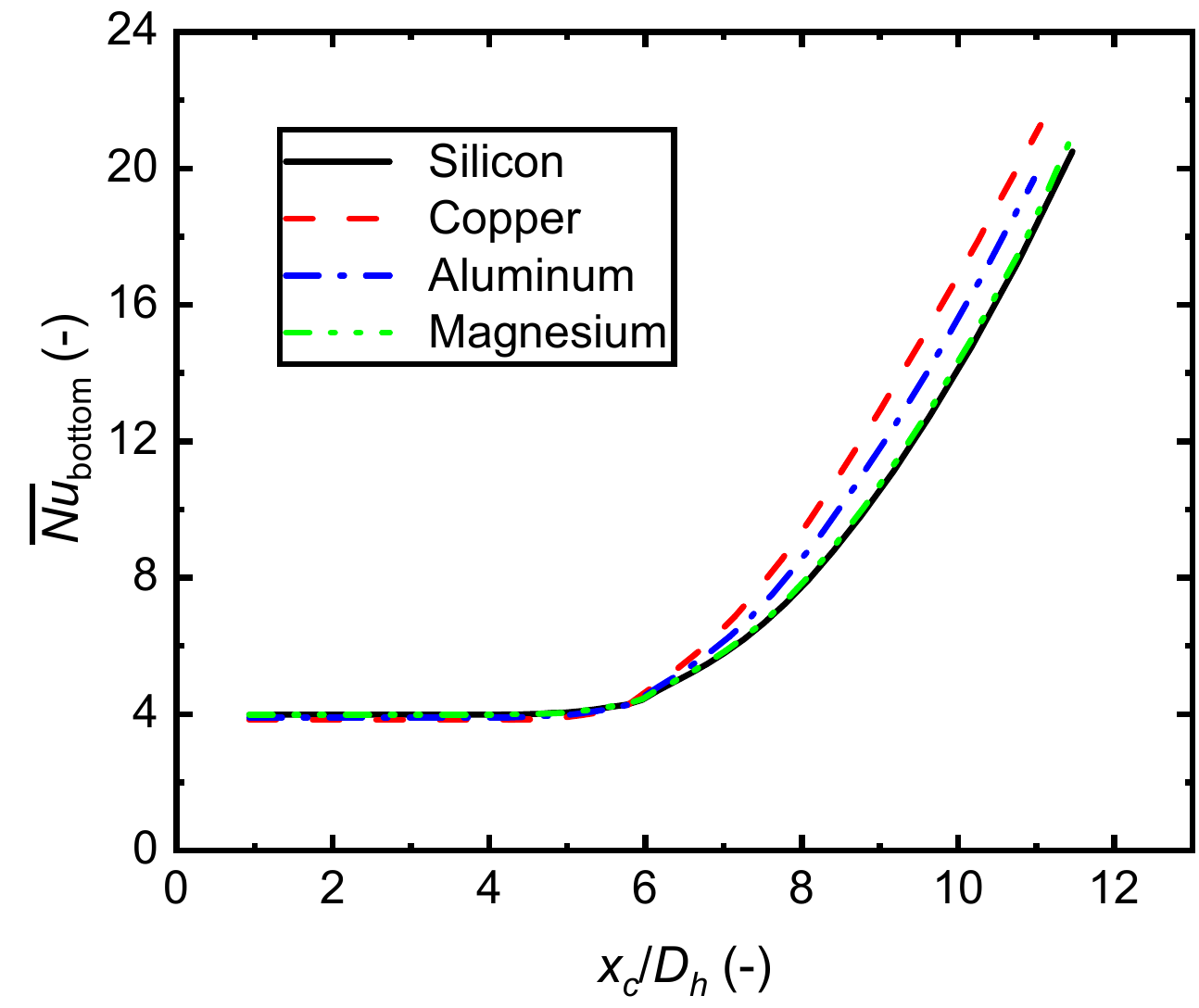}
  \caption{Nusselt number ${{\overline{\Nu}}_{\text{bottom}}}$ plotted versus the dimensionless bubble position $x_c/D_h$ for different solid materials. The bottom wall thickness is $H_b = 40$ $\upmu$m.}\label{fig:13}
\end{figure}

Further, the heat transfer is also quantified by the average Nusselt number at the solid-fluid interface of the bottom wall ${{\overline{\Nu}}_{\text{bottom}}}$. As shown in Figure \ref{fig:13}, the copper microchannel has the highest Nusselt number, which is consistent with the growth of the bubble. This is because of the larger bubble produced in the copper microchannels, which then in turn causes significant perturbation to the velocity field in the microchannel, hence increasing the convection effect. The increase in the flow perturbation in the copper microchannel enhance the heat transfer from the wall to the fluid.

\section{Conclusions}\label{sec:4}
In summary, we explore the conjugate effects during flow boiling in microchannels with the volume of fluid method to capture the bubble growth dynamics. Our results show that the bubble growth and the heat transfer performance are remarkably influenced by the thickness and the material of the channel wall even the heat flux is the same. With increasing the bottom wall thickness, the growth rate of the bubble size increases because of the heat conduction in the solid wall along the channel direction. The solid wall has a higher thermal conductivity than the fluid, hence it can spread heat along the channel direction, increase the wall temperature in the upstream, and accelerate the vaporization of liquid and the growth of the bubble size. The increased bubble size also increases the perturbation to the flow field, and enhances the thermal convection between the fluid and the wall. For different solid materials, the high-thermal-diffusivity material possesses a higher heat transfer performance because it can quickly diffuse thermal energy from the heat source to the solid-fluid interface. Among the microchannel materials considered, copper has the best performance of heat transfer. The findings of this study not only unveil the mechanism of bubble growth in boiling microchannels, but also provide guidance for the design of efficient heat sinks in numerous high-heat-flux applications.

\section*{Acknowledgements}
This work was supported by the National Natural Science Foundation of China (Grant nos.\ 51920105010 and 51921004).


\section*{References}
\bibliography{ChtBoiling}
\end{document}